\documentclass[prl,aps,twocolumn,10pt,showpacs,groupedaddress,superscriptaddress,floatfix]{revtex4-1}
\usepackage{amsmath,amsfonts,amssymb,graphics,graphicx,epsfig,color,times}
\usepackage[utf8x]{inputenc}
\usepackage{color}
\usepackage{bbm, dsfont} 
\usepackage{subfigure}
\usepackage{hyperref}
\usepackage{mathrsfs}
\usepackage{verbatim}
\begin{document}
\title{Phase-imprinted multiphoton subradiant states}
\author{H. H. Jen}
\email{sappyjen@gmail.com}
\affiliation{Institute of Physics, Academia Sinica, Taipei 11529, Taiwan}

\date{\today}
\renewcommand{\k}{\mathbf{k}}
\renewcommand{\r}{\mathbf{r}}
\def\bea{\begin{eqnarray}}
\def\eea{\end{eqnarray}}
\def\ba{\begin{array}}
\def\ea{\end{array}}
\def\bdm{\begin{displaymath}}
\def\edm{\end{displaymath}}
\def\red{\color{red}}
\begin{abstract}
We propose to generate the multiphoton subradiant states and investigate their fluorescences in an array of two-level atoms.\ These multiphoton states are created initially from the timed-Dicke states.\ Then we can use either a Zeeman or Stark field gradient pulse to imprint linearly increasing phases on the atoms, and this phase-imprinting process unitarily evolves the system to the multiphoton subradiant states.\ The fluorescence engages a long-range dipole-dipole interaction which originates from a system-reservoir coupling in the dissipation.\ We locate some of the subradiant multiphoton states from the eigenmodes, and show that an optically thick atomic array is best for the preparation of the state with the most reduced decay rate.\ This phase-imprinting process enables quantum state engineering of the multiphoton subradiant states, and realizes a potential quantum storage of the photonic qubits in the two-level atoms. 
\end{abstract}
\maketitle
{\it Introduction.--}Recent progress on the collective light scattering \cite{Bromley2016, Zhu2016} and subradiant fluorescence measurements \cite{Guerin2016}, and proposals of preparing such subradiant states \cite{Scully2015, Wiegner2015, Jen2016_SR, Jen2016_SR2, Sutherland2016, Bettles2016} opens up a renewed study of strong and collective light-matter interactions \cite{Dicke1954, Stephen1964, Lehmberg1970, Mandel1995, Gross1982} in neutral atoms.\ This cooperative phenomenon of the reduced or prolonged spontaneous emission time results from a long-range dipole-dipole interaction \cite{Stephen1964, Lehmberg1970} in a confined atomic system.\ When the atoms are close enough to each other and become aware of each other's presence through exchanging photons in the dissipation process, the decay rate can be collectively enhanced \cite{Chaneliere2006, Jen2012, Srivathsan2013} along with the associated frequency shift in the emission \cite{Friedberg1973, Gross1982, Scully2009, Rohlsberger2010, Keaveney2012, Meir2014, Pellegrino2014, Jen2015}.\ This collective light-matter interaction can also initiates a subradiant lifetime in nanocavities \cite{Sonnefraud2010} and molecules \cite{McGuyer2015}.\ Furthermore, the atom-atom correlations induced from the dipole-dipole interaction play an important role in determining the emission properties of the cold atoms where the mean-field approach fails \cite{Jennewein2016, Jenkins2016}.\ This pairwise correlation also has an effect on the superradiant laser \cite{Maier2014, Jen2016_SL} in a confined and driven atom-cavity system.

In general a driven light-matter interacting system accesses a total number of $L^N$ states for $N$ atoms with $L$ atomic levels.\ This exponential dependence of number of atoms imposes a limitation on classical computer simulations, where several thousands of atoms are able to be simulated only when, for example, a single excitation is assumed.\ Even such simplification in a reduced Hilbert space of $N$ excited states, a single-photon emission can be superradiant \cite{Scully2006, Eberly2006, Mazets2007, Svidzinsky2008} or subradiant \cite{Scully2015, Jen2016_SR}, which results from the dynamical couplings between these excited states.\ Single-photon superradiance and subradiance, therefore, provide the first step to investigate even more intricate couplings in the multiply-excited Hilbert space.\ For a multiply-excited and subradiant state, it is less explored systematically.\ It is also experimentally difficult to prepare and manipulate these states because of the weak dipole-coupling and intractably large subspace.\ That is, a total of possible states would become \textit{O}$(N^M)$ for $M$ excitations in large $N$ two-level atoms if $M$ $\ll$ $N$.

In this Letter we propose to prepare the phase-imprinted multiphoton subradiant states by applying a Zeeman or Stark field gradient pulse to imprint linearly increasing phases in an array of two-level atoms.\ This multiphoton state can have an extremely low decay rate, which essentially can be stored with a long lifetime \cite{Facchinetti2016}, and is potentially useful in quantum information processing \cite{Hammerer2010} with maximal multiphoton entanglement \cite{Pan2012}.\

{\it Phase-imprinted multiphoton states.--}We consider a setting where a resonant multiphoton is absorbed by two-level atoms with $|g\rangle$ and $|e\rangle$ for the ground and excited states respectively.\ On absorption, the atoms form a symmetrical and multiply-excited state or timed Dicke state \cite{Scully2006},
\bea
|\phi_{C^N_M}^{(M)}\rangle&&=\frac{1}{\sqrt{C^N_M}}\prod_{j=1}^{M}\sum_{\mu_j=\mu_{j-1}+1}^{N-M+j}e^{i\k\cdot\sum_{m=1}^M \r_{\mu_m}}\nonumber\\
&&|e\rangle_{\mu_1}|e\rangle_{\mu_2}\cdots |e\rangle_{\mu_M}|g\rangle^{\otimes(N-M)},\label{sym}
\eea
where $\mu_0$ $=$ $0$.\ We define $\k\cdot\sum_{m=1}^M \r_{\mu_m}$ $\equiv$ $\k\cdot\mathbf{R}_M$ which is the traveling phase in the absorption of the incoming multiphoton with the transition wave vector $\k$ and atomic positions $\r_{\mu_m}$.\ The $M$ atomic excitations are distributed symmetrically among $N$ atoms with $C^N_M$ possible configurations where $C$ denotes the binomial coefficients.\ $C^N_M$ represents the normalization constant for the above collectively excited state and also the number of Hilbert space for the subspace of $M$ excitations in $N$ atoms.\ We note that the initially-prepared state $|\phi_{C^N_M}^{(M)}\rangle$ is not the eigenstate of the system since the dissipation in general involves a resonant long-range dipole-dipole interaction \cite{Lehmberg1970} which we will show later.

\begin{figure}[t]
\centering
\includegraphics[width=8.0cm,height=6cm]{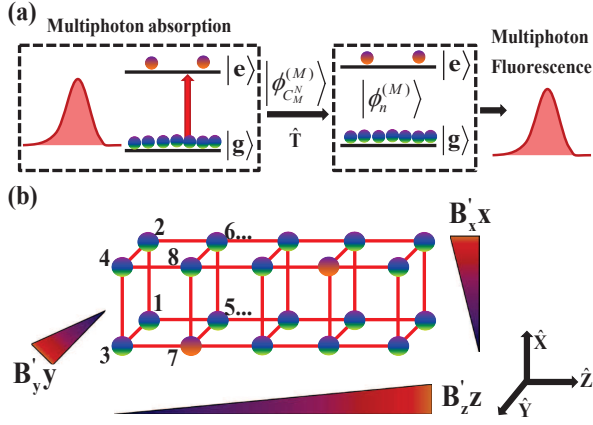}
\caption{(Color online) Schematic phase-imprinted multiphoton subradiant state preparation.\ (a) A multiphoton source is absorbed by the atoms which are excited to the excited states $|e\rangle$ from the ground states $|g\rangle$ (two excitations are shown here for illustration).\ The atomic system is then prepared into the symmetric multiphoton state $|\phi_{C^N_M}^{(M)}\rangle$ which can be unitarily transformed ($\hat{T}$) to $|\phi_{n}^{(M)}\rangle$ by phase-imprinting in (b), where the three-dimensional atomic array acquires linearly increasing phases by three Zeeman magnetic gradient fields $(B_x',B_y',B_z')$.\ $\hat T$ is $e^{iV_B\tau_B}$ with spatial Zeeman energy shift $V_B$ $=$ $\Delta E (\bf{x},\bf{y},\bf{z})$ and interaction time $\tau_B$.}\label{fig1}
\end{figure}

For a concise notation, we define $|\psi_N^{(M)}(\vec{\mu})\rangle$ $\equiv$ $|e\rangle_{\mu_1}|e\rangle_{\mu_2}$ $\cdots$ $|e\rangle_{\mu_M}|g\rangle^{\otimes(N-M)}$ which denote bare state bases of multiple excitations with a configuration $\vec{\mu}$ $\equiv$ $(\mu_1,$ $\mu_2,$ $\cdots,\mu_M)$.\ If there is single excitation ($M$ $=$ $1$), we can construct a complete basis $|\phi_n^{(1)}\rangle$ by imprinting linearly increasing phases on a one-dimensional (1D) \cite{Jen2016_SR} or three-dimensional (3D) atomic array \cite{Jen2016_SR2} of arbitrary $N$ atoms,
\bea
|\phi_n^{(1)}\rangle=\sum_{\mu_1=1}^N \frac{e^{i\k\cdot\mathbf{R}_1}}{\sqrt{N}}e^{i\frac{2n\pi}{N}(\mu_1-1)}|\psi_N^{(1)}(\mu_1)\rangle,
\eea
with $n$ $\in$ [$1$, $C^N_{M=1}$].\ The above includes the symmetrical state of Eq. (\ref{sym}) when $n$ $=$ $N$, and the phase of $2n\pi/N$ can be imprinted by Zeeman or Stark field gradient pulse, as in Fig. \ref{fig1}.\ The orthogonality of the state bases can be satisfied by using De Moivre's formula $\langle\phi_m^{(1)}|\phi_n^{(1)}\rangle$ $=$ $\sum_{\mu_1=1}^N$ $e^{i\frac{2\pi}{N}(\mu_1-1)(m-n)}/N$ $=$ $\delta_{m,n}$.\ For $M$ $\geq$ $2$, the linear phase-imprinted multiphoton states in general cannot construct an orthogonal basis except for $N$ $=$ $3$ and $M$ $=$ $N$.\ We take two excitations, $M$ $=$ $2$, in three equidistant atoms of an array, for example, and 
\bea
|\phi_n^{(2)}\rangle=\sum_{\mu_2=\mu_1+1}^{3}\sum_{\mu_1=1}^{2}\frac{e^{i\k\cdot\mathbf{R}_2}}{\sqrt{3}}e^{i\frac{2n\pi}{C^3_2}[f(\vec{\mu})-1]}|\psi_3^{(2)}(\vec{\mu})\rangle,
\eea
where $f(1,2)$ $=$ $3$, $f(1,3)$ $=$ $4$, $f(2,3)$ $=$ $5$.\ The indices in $f(\vec\mu)$ simply represent a discrete increase of the phases from linear phase-imprinting, that is $f(\mu_1,\mu_2)$ $=$ $\mu_1$ $+$ $\mu_2$.\ This special case of multiphoton states construct a complete Hilbert space of two excitations in a $N$ $=$ $3$ atomic array, which are orthogonal to each other, guaranteed by De Moivre's formula.\ For $M$ $=$ $3$ and $N$ $=$ $3$, the fully-excited state $|\phi_{n=1}^{(3)}\rangle$ becomes symmetrical, as is true for all fully-excited multiphoton states.

On the other hand, for $M$ $\geq$ $2$ and $N$ $>$ $3$, we obtain the phase-imprinted multiphoton states as
\bea
|\phi_n^{(M)}\rangle&&=\prod_{j=1}^{M}\sum_{\mu_j=\mu_{j-1}+1}^{N-M+j}\frac{e^{i\k\cdot\mathbf{R}_M}}{\sqrt{C^N_M}}e^{i\frac{2n\pi}{C^N_M}[f(\vec{\mu})-1]}
|\psi_N^{(M)}(\vec{\mu})\rangle,\label{multiphoton}
\eea
where $n$ $\in$ [$1$, $C^N_M$].\ $f(\vec{\mu})$ $=$ $\sum_{i=1}^M\mu_i$ represents the incremental increases of the imprinted phases in a 3D atomic array as we propose in Fig. \ref{fig1}.\ These $\mu$'s need to be labeled also linearly in a specific order along three axes.\ The order of the axes matters only on the relative gradient strength, and without loss of generality, we choose the labeling order first along $\hat x$, then $\hat y$ and $\hat z$ as shown in Fig. \ref{fig1}.\ The above states $|\phi_n^{(M)}\rangle$ in general cannot construct the orthogonal states in a multiphoton Hilbert space of $M$ excitations since De Moivre's formula does not apply here.\ For example $N$ $=$ $4$ and $M$ $=$ $2$, the bare states $|e_1 g_2 g_3 e_4\rangle$ and $|g_1 e_2 e_3 g_4\rangle$ in Eq. (\ref{multiphoton}) acquire the same phase of $e^{i4n\pi/3}$, which violates the $N$-periodic requirement of De Moivre's formula.\ Nevertheless, our scheme to imprint the linear phases on the atoms can still access the super- and subradiant modes of the system.

{\it Pairwise and long-range dipole-dipole interactions.--}The theoretical analysis for the fluorescence of these subradiant states is based on the Lindblad forms of the dissipation.\ Here we consider a general spontaneous emission process that involves long-range dipole-dipole interaction \cite{Lehmberg1970, Jen2016_SR}.\ This interaction originates from the absorption and reabsorption events in the commonly mediating quantized bosonic field.\ For an arbitrary quantum operator $\hat{Q}$, the Heisenberg equation reads 
\bea
\frac{d\hat{Q}}{dt} = -i\sum_{\mu\neq\nu}^N\sum_{\nu=1}^N G_{\mu\nu}[\hat{Q},\hat{\sigma}_\mu^+\hat{\sigma}_\nu^-] + \mathcal{L}_s[\hat{Q}],\label{Q}
\eea
where the Lindblad form for the spontaneous emission is
\bea
\mathcal{L}_s[\hat{Q}]&=&-\sum_{\mu,\nu=1}^N\frac{F_{\mu\nu}}{2}\left(\hat{\sigma}_\mu^+\hat{\sigma}_\nu^-\hat{Q}+\hat{Q}\hat{\sigma}_\mu^+\hat{\sigma}_\nu^- -2\hat{\sigma}_\mu^+\hat{Q}\hat{\sigma}_\nu^-\right).\nonumber\\
\eea
The annihilation (creation) operator is $\hat{\sigma}_\mu^-$ ($\hat{\sigma}_\mu^+$) where $\hat{\sigma}_\mu^-$ $\equiv$ $|g\rangle_\mu\langle e|$ and $\hat{\sigma}_\mu^-$ $\equiv$ $(\hat{\sigma}_\mu^+)^\dag$.\ The pairwise frequency shift $G_{\mu\nu}$ and decay rate $F_{\mu\nu}$ are \cite{Lehmberg1970}
\bea
F_{\mu\nu}(\xi)&\equiv&
\frac{3\Gamma}{2}\bigg\{\left[1-(\hat{d}\cdot\hat{r}_{\mu\nu})^2\right]\frac{\sin\xi}{\xi}\nonumber\\
&+&\left[1-3(\hat{d}\cdot\hat{r}_{\mu\nu})^2\right]\left(\frac{\cos\xi}{\xi^2}-\frac{\sin\xi}{\xi^3}\right)\bigg\},\label{F}\\
G_{\mu\nu}(\xi)&\equiv&\frac{3\Gamma}{4}\bigg\{-\Big[1-(\hat{d}\cdot\hat{r}_{\mu\nu})^2\Big]\frac{\cos\xi}{\xi}\nonumber\\
&+&\Big[1-3(\hat{d}\cdot\hat{r}_{\mu\nu})^2\Big]
\left(\frac{\sin\xi}{\xi^2}+\frac{\cos\xi}{\xi^3}\right)\bigg\}\label{G}, 
\eea
where $\Gamma$ is the single-particle decay rate of the excited state, $\xi$ $=$ $|\mathbf{k}| r_{\mu\nu}$, and $r_{\mu\nu}$ $=$ $|\mathbf{r}_\mu-\mathbf{r}_\nu|$.\

We then turn to the Schr\"{o}dinger equations which can be projected from the above Lindblad form, to study the time evolution of the phase-imprinted multiphoton states.\ First we define the state of the system with $M$ excitations in Schr\"{o}dinger picture as 
\bea
|\Psi(t)\rangle =\sum_{n=1}^{C^N_M} c_n(t)e^{i\k\cdot\mathbf{R}_M}|\psi^{(M)}_N(\vec{\mu}^n)\rangle,\nonumber
\eea
where $\vec\mu^n$ $\equiv$ $(\mu_1,\mu_2,...,\mu_M)^n$ with $n$ here denoting a specific order.\ Each $n$ requires $\mu_j$ $\in$ $[\mu_{j-1}+1,N-M+j]$ with $\mu_j$ $<$ $\mu_{j+1}$.\ It increases first with $\mu_M$ $=$ $\mu_{M-1}$ $+$ $1$ up to $N$ while fixes the other $\mu_j$'s.\ Next it keeps increasing with an increment in $\mu_{M-1}$ $=$ $\mu_{M-2}$ $+$ $2$ along with $\mu_M$ $=$ $\mu_{M-1}$ $+$ $1$ up to $N$, until $\mu_1$ reaches ($N$ $-$ $M$ $+$ $1$).\ Taking an example of $N$ $=$ $4$ and $M$ $=$ $3$, the bare states $|\psi^{(M)}_N(\vec{\mu}^{n})\rangle$ with $n$ $=$ $1$\textendash$4$ are ordered as $|e_{1}e_{2}e_3g_4\rangle$, $|e_{1}e_{2}g_3e_4\rangle$, $|e_{1}g_{2}e_3e_4\rangle$, and $|g_{1}e_{2}e_3e_4\rangle$.\ Next we obtain the coupled equations of the probability amplitudes according to Eq. (\ref{Q}),
\bea
\dot{c}_n(t)=\sum_{m=1}^{C^N_M} A_{nm}c_m(t),
\eea
where $A_{nm}$ couples all $C^N_M$ states in the subspace.\ The matrix elements are $A_{nn}$ $=$ $-\frac{M\Gamma}{2}$, and $A_{n,m\neq n}$ $=$ $(-F_{s_1s_2}/2+iG_{s_1s_2})e^{-i\k\cdot(\r_{s_1}-\r_{s_2})}$ where $(s_1,s_2)$ can be obtained from a sorting function $S(n,m)$.\ We use $S$ to sort out two numbers, $s_1$ and $s_2$, after comparing the $n$th and $m$th bare states of $|\psi^{(M)}_N(\vec{\mu})\rangle$, which correspond to exactly one different excited atomic index respectively in these bare states.\ Again for the example of $N$ $=$ $4$ and $M$ $=$ $2$, to determine $A_{1,2}$, $S(1,2)$ $=$ $S(|e_{1}e_{2}g_3g_4\rangle,|e_{1}g_{2}e_3g_4\rangle)$ gives $(2,3)$, showing that a dipole-dipole interaction lowers the third atomic excited state while raising the second atom to the excited one.\ This reflects the nature of pairwise couplings in the multiphoton state bases, and thus a matrix $\hat{A}$ represents the dynamically couplings between every two relevant states in $|\psi^{(M)}_N(\vec{\mu})\rangle$.\ If $(s_1,s_2)$ gives $(0,0)$, it leaves a null $A_{n,m\neq n}$ when more than one distinct atomic indices appear in the $n$th or $m$th bare states.\ Using the same example, $A_{1,6}$ ($A_{2,5}$) $=$ $0$, since there is no dipole-dipole coupling between the states $|e_{1}e_{2}g_3g_4\rangle$ and $|g_{1}g_{2}e_3e_4\rangle$ ($|e_{1}g_{2}e_3g_4\rangle$ and $|g_{1}e_{2}g_3e_4\rangle$).\

We can apply the similarity transformation to diagonalize $\hat{A}$ in terms of the eigenvalues $\lambda_l$ and eigenvectors $\hat{U}$, such that the bare state probability amplitude can be solved as
\bea
c_n(t)=\sum_{m,l=1}^{C^N_M} U_{n l}e^{\lambda_l t}U^{-1}_{l m}c_m(t=0),
\eea
where the initial condition, $c_m(0)$ $=$ $1/\sqrt{C^N_M}$, is for the symmetrical Dicke state on absorption of $M$ photons.\ When we evolve the atoms to one of the phase-imprinted multiphoton states $|\phi_n^{(M)}\rangle$ via a unitary transformation $\hat T$ in Fig. \ref{fig1}, the state vector $|\Psi(t)\rangle$ becomes $d_{n}(t)|\phi_{n}^{(M)}\rangle$, and $c_m(0)$ becomes $e^{i2n\pi[f(\vec{\mu})-1]/C^N_M}/\sqrt{C^N_M}$.\ Using the relation of $d_n(t)$ $=$ $\sum_{m=1}^{C^N_M}$ $c_m(t)$ $e^{-i2n\pi[f(\vec{\mu})-1]/C^N_M}/\sqrt{C^N_M}$, where $f(\vec{\mu})$ has an intrinsic dependence of $m$ and follows the same order of the bare states expressing $\hat A$, eventually we obtain the time evolution of the phase-imprinted multiphoton state,
\bea
d_n(t)=\sum_{l=1}^{C^N_M} v_l(n) e^{\lambda_l t}w_l(n),
\eea
where 
\bea
v_l(n)&\equiv&\sum_{m=1}^{C^N_M} \frac{e^{-i2n\pi[f(\vec{\mu})-1]/C^N_M}}{\sqrt{C^N_M}}U_{m l},\\
w_l(n)&\equiv&\sum_{m=1}^{C^N_M} U^{-1}_{l m}\frac{e^{i2n\pi[f(\vec{\mu})-1]/C^N_M}}{\sqrt{C^N_M}}.
\eea
Here $v_l(n)$ is the inner product of $n$th phase-imprinted multiphoton state and $l$th eigenvector in $\hat{U}$, and therefore, $|v_n(m)|^2$ indicates the fidelity to the eigenstate.\ We also define a normalized weighting of $wt(l)$ $\equiv$ $|v_l(n)w_l(n)|^2$ as a measure of the contribution for a specific $\lambda_l$ which determines the time evolution of the phase-imprinted multiphoton states.

\begin{figure}[b]
\centering
\includegraphics[width=8.5cm,height=4.5cm]{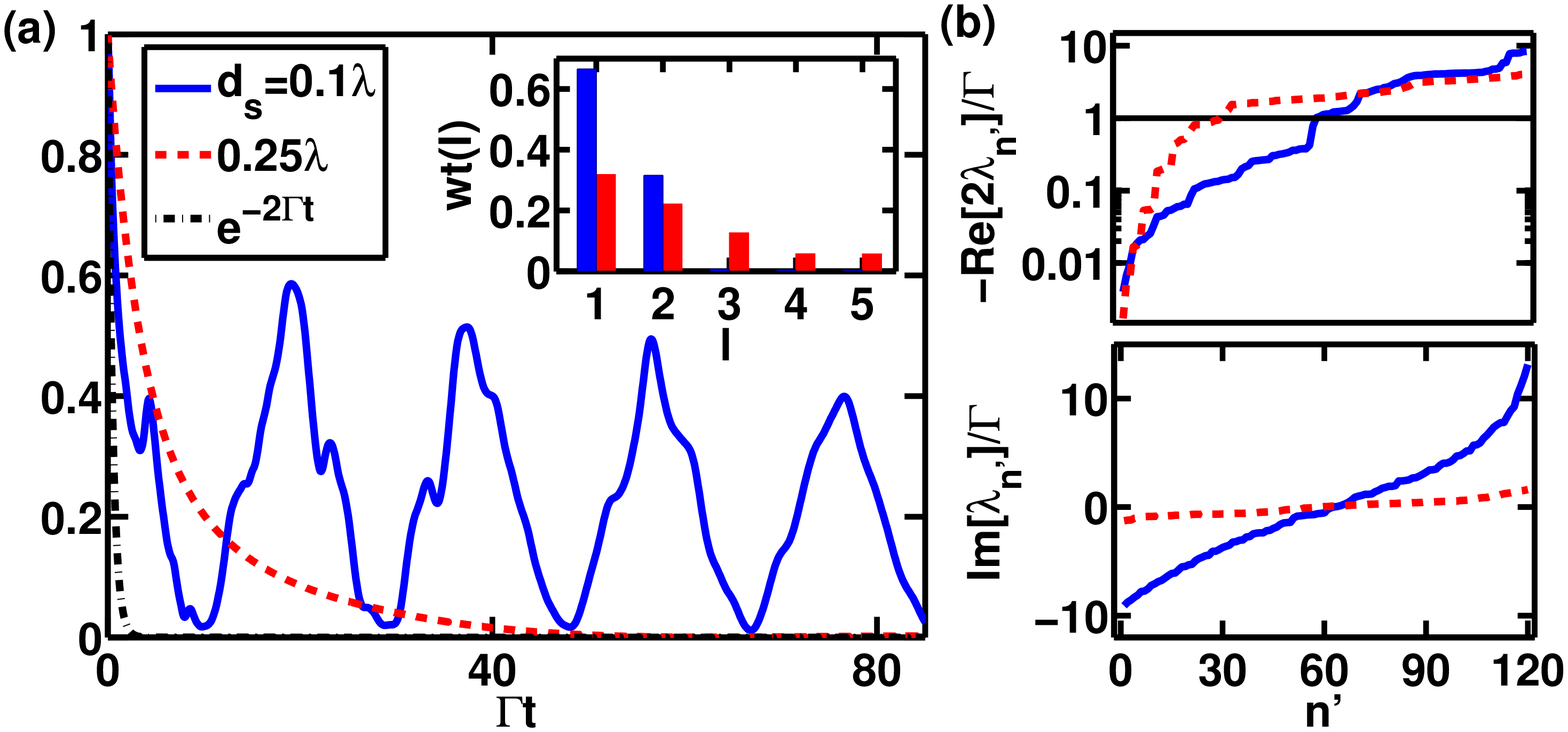}
\caption{(Color online) Time evolutions of phase-imprinted two-photon subradiant states for different atomic separations.\ (a) For 1D atomic array of $N$ $=$ $16$ and $d_s$ $=$ $0.1$ (solid) and $0.25\lambda$ (dash), we show the oscillatory time evolutions of the two-photon subradiant state with $n$ $=$ $45$ comparing the intrinsic decay $e^{-2\Gamma t}$ (dash-dot).\ The inset shows five most significant weightings of $|\phi_n^{(2)}\rangle$ on the eigenstates $|\phi'_{n'}\rangle$, where left (right) bar corresponds to $d_s$ $=$ $0.1\lambda$ ($0.25\lambda$).\ (b) The ascending order of the real and imaginary parts of the eigenvalues in a total of $C^{16}_2$ $=$ $120$ states.\ A horizontal line guides the eye for a natural decay constant of $-$Re$[2\lambda_{n'}]/\Gamma$ $=$ $1$.}\label{fig2}
\end{figure}

{\it Fluorescence of multiphoton subradiant states.--}Time evolutions of the the phase-imprinted multiphoton states correspond to the fluorescence observed in experiments.\ In Fig. \ref{fig2}, we show the time evolutions of the two-photon subradiant states in an atomic array of sixteen atoms, and consider $\hat x$-polarized excitations.\ As a comparison, we choose the phase-imprinted states of $n$ $=$ $45$ in two different lattice spacings of $0.1$ and $0.25\lambda$.\ Both cases demonstrate subradiant decays comparing the intrinsic spontaneous emission of two excited atoms $e^{-2\Gamma t}$.\ A decayed Rabi oscillation in Fig. \ref{fig2}(a) also appears in the single-photon subradiant state \cite{Jen2016_SR}, and its beating frequency can be determined by the difference of the frequency shifts in the eigenstates.\ Five most significant weightings [$wt(l)$] on the eigenstates are plotted in the inset of Fig. \ref{fig2}(a), where $l$ $=$ $1$ and $2$ at $d_s$ $=$ $0.1\lambda$ correspond to $2\lambda_{n'}/\Gamma$ $=$ $0.004$ $-$ $i18.19$ and $0.0068$ $-$ $i17.53$ respectively.\ The difference of $0.33\Gamma$ in the frequency shifts thus gives the oscillation period $T$ $=$ $2\pi/0.33\Gamma$, or $\Gamma T$ $\sim$ $20$.\ Taking rubidium atoms as an example, the envelope of this oscillation can be fitted to around $5$ $\mu$s, showing $\sim$ $200$ times longer than the intrinsic decay time $1/\Gamma$ $\sim$ $26$ ns.\ Several small humps and peaks are present due to the other minor weightings (almost negligible) of $wt(l\neq 1,2)$.\ In contrast to this oscillatory feature, the phase-imprinted state at $d_s$ $=$ $0.25\lambda$ simply decays with a subradiant rate.\ This is due to the extended distribution of the weightings, which smooths out various beating frequencies in these eigenvalues.\ As expected, a shorter lattice spacing allows to have more enhanced superradiant modes and frequency shifts, which can be seen in Fig. \ref{fig2}(b) at $n'$ $\gtrsim$ $90$.

\begin{figure}[t]
\centering
\includegraphics[width=8.5cm,height=4.5cm]{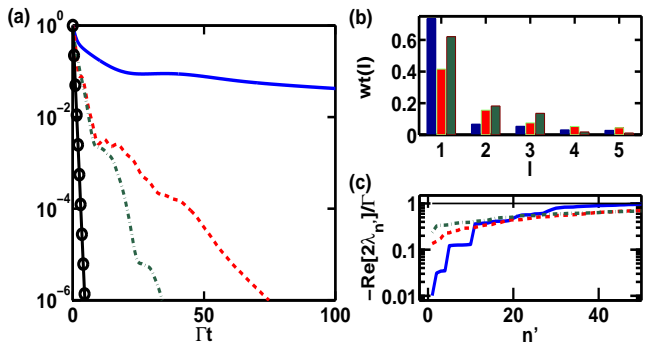}
\caption{(Color online) Time evolutions of phase-imprinted three-photon subradiant states for different atomic geometries.\ (a) Different atomic configurations for $N$ $=$ $16$ and $d_s$ $=$ $0.25\lambda$ involve $N_x\times N_y\times N_z$ $=$ $1\times 1\times 16$ (solid), $2\times 2\times 4$ (dash), and $4\times 4\times 1$ (dash-dot).\ The time evolutions of the subradiant states are demonstrated in a logarithmic scale with $n$ $=$ $135$, $100$, and $70$.\ These $n$'s are chosen for a relatively large $l$ $=$ $1$ weighting among all, respectively corresponding to the lowest decay rate in the eigenvalues.\ The intrinsic decay of $e^{-3\Gamma t}$ is denoted ($\circ$) as a comparison.\ (b) Five most significant weightings are shown for these geometries respectively as the left, middle, and right bars.\ (c) Subradiant eigenvalues (real parts) out of a total of $C^{16}_3$ $=$ $560$ ones are shown in a logarithmic scale.}\label{fig3}
\end{figure}

For even more photons in different geometries, we investigate the phase-imprinted three-photon subradiant states in Fig. \ref{fig3}, generated from an array, rectangular cuboid, to a square.\ This shows an effective transition from optically thick to optically thin atoms.\ The lifetimes of these subradiant states can be estimated from the time evolutions in Fig. \ref{fig3}(a).\ The atomic array shows either $120$ or $200$ times the lifetime [$1/(3\Gamma)$] of the intrinsic spontaneous decay $e^{-3\Gamma t}$, with a fitting to the intercept at the origin or not.\ This fitting deviation is due to an initial rapid drop and a following flattened decay, indicating an non-exponential behavior in the subradiance.\ For the optically thin shapes, fitting an exponential decay is sufficiently well, and their lifetimes are only about four to five times longer, where the square lattice shows less.\ The weightings on the eigenvalues in Fig. \ref{fig3}(b) rather spread out.\ Therefore, this limits the sole access to the lowest possible decay constant in Fig. \ref{fig3}(c), where the lowest decay rates can be as reduced as a hundred or ten times smaller respectively for the array or optically thin geometry.\ 

To access even lower subradiant decay rates, we can either increase the total number of atoms or consider an optically thick system (longer atomic chain).\ For $N$ $=$ $27$ and $M$ $=$ $3$, we obtain the lowest decay rates of $4.5\times 10^{-4}$ and $0.04\Gamma$ respectively in the array and the cube.\ For a lattice spacing $d_s$ $\gtrsim$ $0.5\lambda$, the effect of resonant dipole-dipole interaction in the dissipation is not significant, and therefore the fluorescence of the multiphoton states does not distinguish much from the noninteracting ones.\ Due to the limitation of available Zeeman field gradient in several mG$/\mu$m, practically it is not possible to efficiently prepare all the phase-imprinted subradiant states proposed in Eq. (\ref{multiphoton}).\ For a longer atomic array, a more demanding field gradient is required to imprint the phase difference between two adjacent sites.\ For example, preparing the $n$th multiphoton subradiant state requires a phase difference of $2n\pi/C^N_M$.\ However only a fraction of $2\pi$ phase difference is needed to show significant subradiance, as shown in Figs. \ref{fig2} and \ref{fig3} where $n$ $\ll$ $C^N_M$.\ 

Other issue for phase fluctuations can be removed if the stability of atomic chains and field gradients can be sustained.\ In experiment, a few percent level of phase error can be controlled, and therefore the phase-imprinted scheme should be robust enough to prepare the multiphoton subradiant state with a high fidelity \cite{Jen2016_SR}.\ On the other hand, the low efficiency due to the short lifetime in the initialized timed-Dicke state can be restored by heralded multiphoton source and post selection.\ The phase-imprinted scheme promises a dynamical control over the subradiant states, which can be implemented as quantum memory of a multiphoton source.\ The radiation properties can be also engineered via manipulating atomic geometries and lattice spacings.\ We note that a complexity of larger $N$ and $M$ arises quite fast already for three atomic excitations in twenty-seven atoms.\ In such case, a dimension of $2925$ states appears in the eigenvalue algorithm, comparing several hundreds of states in Figs. \ref{fig2} and \ref{fig3}.\ We estimate the time consumed in simulating the case for $N$ $=$ $20$ and $M$ $=$ $10$ to longer than seven days, using a typical Intel(R) Xeon CPU E5.\ Therefore even for such a few-body system with an order of ten atomic excitations, the whole Hilbert space has challenged the capability of present computations.\ 

In conclusion, we propose a scheme to prepare phase-imprinted multiphoton subradiant states in two-level atomic arrays.\ The fluorescences of these states decay with significantly reduced rates and can be oscillatory with a beating frequency determined by the difference of the frequency shifts.\ The optically thick atomic system with the appropriate imprinted phases is best for the preparation of the subradiant state with the longest lifetime, making a promising two-level atomic quantum memory of multiphoton.\ Future study can also apply this phase-imprinted scheme to a ring-shaped atomic array or the atoms distributed on the side of a cylinder.\ Using light excitations with $m$ orbital angular momentum \cite{Arnaut2000, Mair2001, Molina2007, Dada2011, Fickler2012} automatically imprints a phase of $e^{i m\phi}$ on the atoms, and essentially the phase-imprinted multiphoton states proposed here can be the candidates for quantum storage of orbital angular momentum photonic qubits \cite{Nicolas2014, Ding2015, Zhou2015}.

{\it Acknowledgements.--} This work is supported by the Ministry of Science and Technology (MOST), Taiwan, under the Grant No. MOST-104-2112-M-001-006-MY3.\ We are grateful for stimulating discussions with Pye-Ton How on a sorting algorithm in assigning multiphoton coupling matrix elements, and with M.-S. Chang and Y.-C Chen on a potential experimental implementation in the ring-shaped atomic arrays.

\end{document}